# Pathways of early evolution from the perspectives of a riboreplisome – the ultimate RNA machine of life

Alice Cleynen[1,2], Archa H. Fox[3], Nikolay E. Shirokikh[3]

[1]University of Montpellier, CNRS, Montpellier, France
[2]IRL FAMSI, CNRS international lab, Australian National University, Canberra, Australia and
[3]Australian Centre for RNA Therapeutics in Cancer, School of Human Sciences, University of Western Australia, Perth, Australia

**Abstract**

Translation of the genetic code into proteins is universal across life, with ribosomes serving as the ancient molecular machines enabling this information transformation. We propose that early replication and protein biosynthesis were inseparable functions performed by a single ancient RNA molecule: the riboreplisome. This hypothesis addresses fundamental challenges facing RNA world, protein world, and RNA-protein co-evolution theories – particularly the problem of non-Darwinian starting points and the interdependence of replication and translation machineries. We present a 12-step evolutionary pathway from simple RNA replicase to proto-ribosome, supported by quantitative modelling demonstrating that only two steps require non-Darwinian transitions while the remaining ten proceed through standard selection. The single-molecule riboreplisome architecture provides an approximately 6-fold evolutionary advantage over multi-component "RNA soup" scenarios, principally by maintaining genetic linkage and retaining catalytic products close to the replicating genome (resisting diffusion loss); structural resistance to parasitic "cheater" molecules follows as a corollary. The riboreplisome framework explains the origins of ribosomal, transfer, and messenger RNA as derivatives of a single ancestral molecule, while providing a driving force for cellular compartmentalisation. Although the original riboreplisome is likely lost to deep time, molecular remnants may persist in extant biological RNA. Identification or reconstruction of riboreplisome-like molecules could advance both origin-of-life research and synthetic biology applications.

## Introduction

Cellular evolution has generated immensely complex molecular information systems built upon RNA activity (Delihas, 2006; Higgs and Lehman, 2015). RNA programs diverse cellular functions across organisms ranging from microbes to mammals, playing indispensable roles in development and differentiation (Mattick and Amaral, 2022). Yet the pre-cellular "chemical" or "abiotic" origins of life-related molecular functions remain deeply enigmatic (Cech, 2012; Joyce, 1989; Lazcano and Miller, 1996). While the precise chemophysical environment of early evolution may never be known with certainty, modern RNA biology provides crucial clues: the abundance of energy-consuming RNA chaperones that maintain RNA in functional conformations (Lüking et al., 1998; Rajkowitsch et al., 2007) suggests primordial conditions may have differed substantially from present-day cellular environments.

The ribosome presents a unique puzzle for origin-of-life research. Unlike most enzymatic systems, it operates with extraordinary specificity on remarkably diverse substrates – including the effectively sequence-nonspecific mRNA. The ribosome coordinates over 100 transfer RNAs with multiple aminoacyl-tRNA synthetases to achieve programmable protein synthesis (Gomez and Ibba, 2020; Ibba and Söll, 2000; Schimmel, 1987). How could such an interconnected, self-dependent system have emerged? This question exemplifies the challenge identified in major evolutionary transitions (Szathmáry, 2015): explaining the origin of new levels of biological organisation without invoking implausibly complex starting points. Traditional approaches – the RNA world (Gilbert, 1986), protein world (Kurland, 2010), and RNA-protein co-evolution hypotheses (Bowman et al., 2015) – each face variants of this “chicken-and-egg” problem. The ribosome requires proteins for efficient function, yet proteins require ribosomes for their synthesis. Following Eigen's framework for self-organisation of biological macromolecules (Eigen, 1971) and Kauffman's principles of autocatalytic sets (Kauffman, 1993), we propose that this apparent paradox dissolves when replication and translation are recognised as originally unified functions of a single molecular entity.

Here we present the riboreplisome hypothesis: that early replication and proto-ribosomal functions were performed by the same ancient RNA molecule. This framework provides an uninterrupted evolutionary thread from chemistry to biology, consistent with Jacob's principle that evolution proceeds by tinkering with existing structures rather than designing *de novo* (Jacob, 1977). We outline a 12-step pathway from simple RNA replicase to compartmentalised proto-cells, supported by quantitative modelling demonstrating that the single-molecule riboreplisome architecture provides fundamental evolutionary advantages over multi-component scenarios. Critically, only two of twelve steps require non-Darwinian transitions – the remainder proceed through standard selection, with transition probabilities five orders of magnitude higher than the initial bottleneck events. The riboreplisome hypothesis thus offers a parsimonious resolution to long-standing challenges in understanding the emergence of the genetic code and its decoding machinery.

**Early link of replication with the mechanisms of replication enhancement**

Sequential improvement through generation of close copies defines evolution itself. Consequently, the emergence of any proto-ribosomal structure cannot be separated from RNA replication – the protoribosome must have been either part of the replicase or co-evolved to enhance replication efficiency. This reasoning leads inevitably to the concept of a “riboreplisome”: an RNA molecule combining replication and proto-translational functions that can lead to the “expression of genes”.

Experimental evidence strongly supports the plausibility of RNA-based replication. Synthetic evolution experiments have generated RNA ligase ribozymes capable of template-directed polymerisation (Attwater et al., 2018; Ekland and Bartel, 1996; Johnston et al., 2001; Tjhung et al., 2020; Wochner et al., 2011). Critically, these ligases were selected from random sequence pools rather than derived from natural ribozymes, demonstrating that functional replicases can emerge *de novo* (Bartel and Szostak, 1993; Ekland et al., 1995). Recent work has further shown that directed evolution can rapidly generate diverse ligase reactivities – including a single point mutation that toggles between linear ligation and branching – highlighting the evolutionary flexibility inherent in RNA catalysis (Biswas and DasGupta, 2026). Recent cryo-EM structural determination of an RNA polymerase ribozyme has revealed how such replicases achieve their function: the triplet polymerase ribozyme operates as an RNA heterodimer with catalytic and scaffolding subunits connected by kissing-loop interactions, converging on domain architectures remarkably similar to protein polymerases despite independent evolutionary origins (McRae et al., 2024). Most strikingly, Gianni et al. recently discovered QT45, a 45-nucleotide polymerase ribozyme evolved *de novo* from random sequence pools that can synthesise both its complementary strand and a copy of itself using trinucleotide building blocks – the two key reactions required for self-replication (Gianni et al., 2026). This demonstrates that RNA polymerase activity can be encoded in motifs far smaller and simpler than the class I polymerases (>150 nucleotides), substantially increasing the plausibility of spontaneous replicase emergence. Given this experimental foundation, what constraints would shape early replicase architecture?

Several replication topologies merit consideration (**Figure 1**). Complete single-molecule self-replication poses a fundamental problem: the polymerase domain must unfold during its own replication (**Figure 1a**). Possible solutions include dual-domain architectures that hand substrates between replication centres (**Figure 1b**), or multi-copy arrangements tolerant of incomplete replication through slippage-like mechanisms (**Figure 1c**) (Canceill et al., 1999; Zhang et al., 2020). Alternatively, replication may occur exclusively *in trans* between molecular partners (**Figure 1d,e**).

*In cis* replication offers built-in resistance to dilution but risks self-inactivation through duplex formation with the nascent negative strand. While environmental cycling (*e.g.*, temperature fluctuations) could resolve duplex-induced shutdown, dependence on external cycling represents an evolutionary disadvantage. *In trans* replication simplifies the chemistry but requires mechanisms for partner capture and retention. Both variants likely coexisted during early evolution.

One limiting, maximally-linked topology is an inverted-repeat architecture, in which replicase function is retained on both positive and negative strands (**Figure 1e**). Such arrangements, analogous to inverted terminal repeats in modern DNA viruses, would be fully self-contained and dilution-resistant. We note, however, that this is the most demanding case: the reverse complements of characterised polymerase ribozymes are generally non-functional, and a longer inverted repeat also pays a replication-length cost relative to a shorter, asymmetric replicase. We therefore treat it as illustrative rather than required – the genetic-linkage advantage that drives the rest of this work is provided equally by the simpler *in cis* and partner-dependent topologies (**Figure 1a–d**). The decisive currency is not per-cycle copying speed but heritable retention of beneficial variation together with local retention of catalytic products: in any linked, *in cis* arrangement, catalytic improvements are co-replicated with the genome and products are generated next to the molecule that encodes them, whereas a faster free replicase copying separate templates forfeits this genotype–phenotype association (quantified in Model 3, below). Even in this scenario, asymmetric activity between strands would establish a generational cycle and, effectively, the first “gene” subject to expression control. These molecular dynamics bear striking resemblance to biological phenomena including sexual reproduction, hermaphroditism, and alternation of generations.

The structural feasibility of cooperative, multi-subunit RNA replicase architectures is supported by recent structural work. The cryo-EM structure of an evolved RNA polymerase ribozyme demonstrates that functional replicases can operate as multi-subunit complexes, with kissing-loop interactions providing the intermolecular contacts necessary for catalytic cooperation (McRae et al., 2024; Shechner, 2024). This validates the feasibility of partner-dependent, intermolecular (*in trans*) replicase cooperation (**Figure 1d**). We emphasise, however, that it does not demonstrate catalytic activity of a reverse-complement strand; since the reverse complements of characterised polymerase ribozymes are generally non-functional, the fully inverted-repeat arrangement (**Figure 1e**) remains a more speculative, limiting case. The hypothesis does not depend on it: what is essential is only that replication function and genome reside on the same molecule (*in cis*), so that copying acts on “self”, a condition met equally by the *in cis* and partner-dependent topologies of **Figure 1a-d**.

Since *in trans* replication appears more chemically tractable, selection would favour mechanisms for binding replication partners, which can be most readily achieved through nucleotide complementarity. Importantly, such partner retention does not require sequence-specific template discrimination by the replicase: a promiscuous replicase still copies its own linked or complementary partner preferentially, simply because that template is held at the highest local concentration and is physically retained. Self-preference is therefore topological rather than sequence-based, which is why the scheme does not demand the strict specificity that characterised polymerase ribozymes lack, and why a free parasite, which is neither linked nor locally concentrated, gains no lasting advantage from being copied *in trans*. Remarkably, *in vitro* evolution experiments demonstrate that such complementarity-based interactions emerge rapidly in self-replicating RNA systems and bear notable resemblance to Shine-Dalgarno:anti-Shine-Dalgarno interactions used in modern translation initiation (Wochner et al., 2011) . This convergence hints at deep mechanistic connections between primordial replication and the emergence of ribosomal function.

Eigen’s theoretical framework for self-organising macromolecular systems (Eigen, 1971) provides quantitative grounding for these considerations. His analysis established that replicating systems face an “error threshold”, a maximum genome length sustainable at a given replication fidelity. For primitive RNA replicases with error rates of ~$10^{-2}$ per nucleotide, this threshold limits functional sequences to roughly 100 nucleotides. The riboreplisome architecture, by coupling replication enhancement directly to the replicating molecule, provides a mechanism for progressively raising this threshold rather than violating it. The error threshold is a length–fidelity relationship (the maximum sustainable genome scales inversely with the per-nucleotide error rate), so a system that begins small (well within the limit) can grow only as fast as its fidelity improves. Because fidelity-enhancing variation is part of the replicating molecule, every such improvement is immediately heritable and immediately raises the sustainable length: size and fidelity therefore co-expand, and genome growth along our pathway is gradual and selection-gated. This is an alternative to Eigen’s hypercycle, which instead partitions information across several independently replicating species; the hypercycle evades the threshold at the cost of the coordination, diffusion, and parasite (short-circuiting) problems that the linked, single-molecule riboreplisome avoids by construction. The recent discovery of a 45-nucleotide polymerase ribozyme capable of self-copying (Gianni et al., 2026) demonstrates that functional replicases can exist well within this error threshold constraint, as the short sequence length dramatically reduces the fidelity required for faithful replication.

**Figure 1.** Replication topologies and the riboreplisome solution. (**Left panel**) Replication challenges. Possible topologies of early RNA ribozyme-based replication, each presenting distinct evolutionary constraints. 5′ ends are marked with filled circles; 3′ ends are indicated as "–OH". **(a)** Single-molecule, single-domain *in cis* self-replication is problematic as the replicase must unfold during its own copying. Inset shows the secondary structure and cryo-EM-derived tertiary structure of the triplet polymerase ribozyme (TPR), an RNA heterodimer comprising catalytic (orange, 5TU) and scaffolding (cyan, t1) subunits connected by kissing-loop interactions ((McRae et al., 2024); PDB: 8T2P). This structure illustrates the architectural complexity achievable by RNA polymerase ribozymes. **(b)** Two-domain self-replication is possible but requires coordinated substrate transfer between replication centres. **(c)** Multiple-copy replicase arrangement is possible but requires mechanisms of copy regeneration, such as repeat slippage. **(d)** *In trans* replication with an inert negative strand intermediate (purple) operates at half capacity since only one strand carries

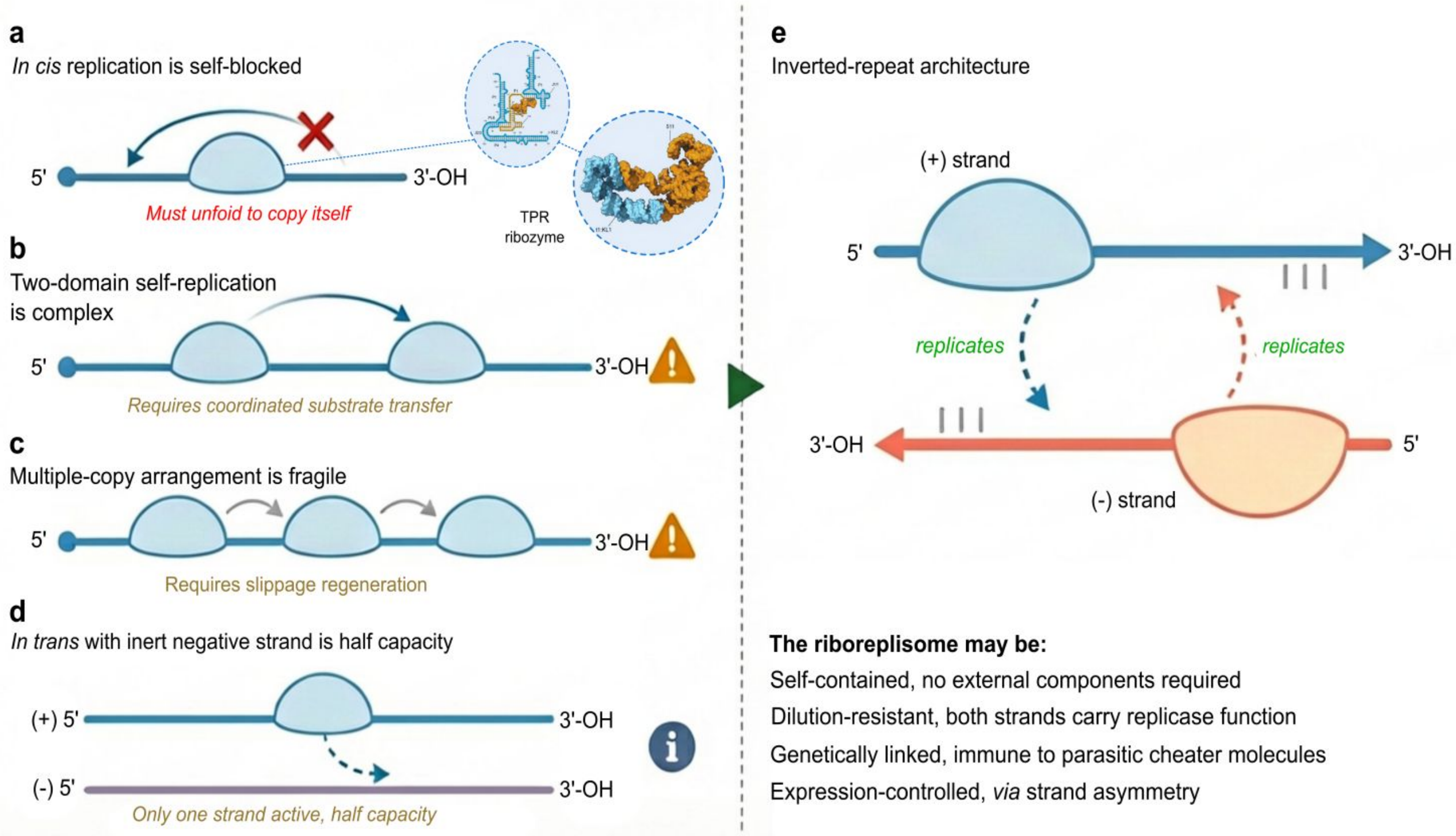

polymerase function. **(Right panel)** The riboreplisome solution. **(e)** Inverted-repeat architecture enables *in trans* replication where both positive (blue) and negative (orange) strands encode functional replicase domains. This topology is self-contained, dilution-resistant, and genetically linked, so that catalytic products and beneficial variation remain associated with the genome; structural resistance to parasitic "cheater" molecules follows as a corollary. Asymmetric activity between strands establishes the first "gene" subject to "expression" control. Key evolutionary advantages are summarised; quantitative modelling demonstrates approximately 6-fold advantage over multi-component scenarios (see Models 1–3, below). This architecture leads to the 12-step evolutionary pathway detailed in **Figures 2-3**.

**Theories of ribosomal organisation and emergence**

Six decades of research have generated substantial evidence regarding ribosome evolution, converging on several key principles directly relevant to the riboreplisome hypothesis. We summarise these findings here. Woese's foundational work established that early translation machinery was far simpler and more error-prone than modern ribosomes, likely involving a limited set of RNA molecules (perhaps just one) with no clear functional distinction between proto-rRNA, proto-tRNA, and proto-mRNA (Woese, 1965; Woese et al., 1966). This prescient insight aligns remarkably with the riboreplisome concept. Subsequent structural and phylogenetic analyses have consistently identified the peptidyl transferase centre (PTC) as the most ancient ribosomal structure, with regions responsible for tRNA interaction showing signs of later co-evolution (Caetano-Anollés, 2002; Petrov et al., 2015, 2014; Smith et al., 2008). Three competing frameworks address ribosome origins: the RNA world hypothesis (Gilbert, 1986), the protein world hypothesis (Kurland, 2010), and RNA-protein co-evolution models (Bowman et al., 2015; Harish and Caetano-Anollés, 2012; Tagami and Li, 2023). While each offers valuable insights, all share a critical limitation: they do not adequately explain how beneficial features could be both generated and selected before the establishment of genetic linkage between genotype and phenotype (Bowman et al., 2015). The protein world hypothesis, despite intriguing structural-cladistic support, lacks clear mechanisms for information inheritance. RNA-protein co-evolution models require implausibly complex starting conditions. Even the RNA world hypothesis struggles to explain the transition from random RNA catalysis to heritable, selectable functions. Several structural findings are particularly relevant to the riboreplisome framework. First, the "fossil onion" model demonstrates that the ribosome can be conceptually "peeled" to reveal progressively more ancient structures centred on the PTC (Hsiao et al., 2009). The RNA fragments constituting this ancient core are notably compact, approximately 100 nucleotides or less, consistent with Eigen's error threshold constraints (Smith et al., 2008). Second, rRNA folding studies reveal no strict requirement for sequential assembly; multiple concurrent folding pathways exist, with the most ancient PTC regions paradoxically folding last (Davis et al., 2016; Dong et al., 2023). This flexibility suggests that early proto-rRNA could fold acceptably without elaborate protein assistance. Third, and most critically, recent experiments demonstrate that the ribosome can function as a single covalently-linked RNA molecule. "Tethered" ribosomes, with subunits connected *via* an RNA linker, retain sufficient activity to support cellular function (Aleksashin et al., 2019; Schmied et al., 2018). While less efficient than wild-type ribosomes likely due to suboptimal folding rather than fundamental catalytic limitations, tethered ribosomes prove that subunit division is not essential for translation. This finding directly supports the possibility of an ancestral single-molecule riboreplisome. The reduced efficiency of tethered ribosomes may reflect evolutionary "channelling": once subunit division was established, optimisation proceeded so rapidly that reverting to a single-molecule architecture became competitively impossible. Complementing the tethered ribosome work, minimal PTC constructs demonstrate that short RNA fragments (~67 nucleotides) can catalyse peptide bond formation, albeit at rates ~200-fold lower than intact ribosomes (Bose et al., 2022). This represents remarkable efficiency given the dramatic size reduction, and establishes the plausibility of RNA-only proto-ribosomal catalysis. Factor-free translation experiments further confirm that translation elongation factors EF-Tu and EF-G, while enhancing efficiency, are not strictly required for peptide synthesis (Gavrilova et al., 1976). The genetic code itself provides additional clues. The "descending thermostability" hypothesis proposes that early codons relied on strong G-C interactions, with weaker A-U codons added later as the machinery became more sophisticated (Grosjean and Westhof, 2016; Trifonov, 2000). Initial codes may have been dinucleotide-based, encoding only simple amino acids like glycine, alanine, and proline (Hartman and Smith, 2014). This progressive code expansion is consistent with the stepwise sophistication predicted by the ribोreplisome model. Recent experimental advances have strengthened the case for RNA-peptide co-evolution from the earliest stages. Singh *et al.* demonstrated spontaneous, selective aminoacylation of RNA *via* thioester intermediates under prebiotically plausible conditions, representing the first achievement of this critical reaction in water at neutral pH (Singh et al., 2025). This breakthrough directly supports the riboreplisome hypothesis by showing that self-aminoacylation (Steps 3-4 of our proposed pathway) is chemically feasible. Furthermore, hydrophobic cationic peptides have been shown to enhance RNA polymerase ribozyme activity through accretion-like mechanisms (Li et al., 2022), suggesting that even simple peptides could have provided immediate selective advantages to early riboreplisome variants. Despite these advances, existing theories share a fundamental gap: they do not explain how the ribosome's remarkable interdependence and simultaneously requiring tRNAs, mRNA, and aminoacyl-tRNA synthetases for function could have evolved without an implausibly complex starting point. The riboreplisome hypothesis, developed in the following section, addresses this gap by proposing that these components originated as domains of a single, genetically unified molecule.

**Emergence of the current ribosomal organisation as the possible converging point**

The ribosome is fundamentally an RNA machine: ribosomal RNA self-folds and drives assembly, and the peptidyl transferase centre is a ribozyme essentially devoid of protein. The experimental support reviewed above, including RNA-only PTC catalysis by short fragments, the viability of a single covalently-linked ("tethered") ribosome, factor-free elongation, and ribozyme-catalysed aminoacylation, establishes that an RNA-only proto-ribosome is feasible. The key implication is directional: protein is the evolutionary product of the ribosome, not its precursor, and there may have been no purpose-made proteins when the riboreplisome emerged, or these were made through completely different pathways that became extinct.

Most existing theories propose multi-component co-evolution of proto-tRNAs, proto-rRNAs, and proto-mRNAs. However, this view faces several difficulties. Ribozyme-based replication can achieve reasonable fidelity even after limited evolution (Johnston et al., 2001; Tjhung et al., 2020). Transcription errors, by definition, are not individually heritable and thus not subject to natural selection. Without cell-like compartmentalisation, functional gains cannot be linked to their genetic carriers. Mechanisms for assembling short proto-rRNA fragments into longer chains remain unexplained. These multi-component models contain considerable "pure" Lamarckism, perhaps acceptable for certain functions in complex life systems with *e.g.* established epigenetic mechanisms, but problematic for the earliest genetic evolution.

We combine these observations to propose the riboreplisome: an RNA molecule with features of a complete genetic system. The following 12 steps describe its emergence and early evolution, with only Steps 1-2 requiring non-Darwinian transitions; the immediate selective advantage conferred by each step is summarised in **Table 1** (**Figures 2, 3**).

**(1)** Small oligomeric RNAs spontaneously develop ligation and limited templated polymerisation activity, experimentally demonstrated to be plausible (Attwater et al., 2018; Bartel and Szostak, 1993; Ekland et al., 1995) (**Figure 2a**, left). Biswas and DasGupta recently showed that single point mutations can toggle ribozyme function between linear ligation and branching, highlighting the evolutionary flexibility inherent in such early RNA catalysts (Biswas and DasGupta, 2026). This step likely occurred multiple times without subsequent development before acquisition of additional features.

**(2)** Templated polymerisation becomes more ordered through basepairing-induced interactions with replication partners, using topologies discussed earlier and in **Figure 1** (Tjhung et al., 2020). This forms an RNA ribozyme “replisome” that restricts copying to “self” rather than “others” (**Figure 2a**, right). The emergence of self-recognition establishes a selective framework in which subsequent innovations can be evaluated and retained. From this point, Darwinian natural selection and evolution proceed uninterruptedly.

**(3)** The RNA replisome, capable of interacting with RNA 3' ends for replication, develops ability to modify (“charge”) its own 3' end with different chemistry, including amino acids (Lee et al., 2000). Charging occurs in bind-and-release cycles using environmentally available aminoacyl-AMPs (**Figure 2b**). The recent demonstration of spontaneous RNA aminoacylation *via* thioesters under prebiotic conditions (Singh et al., 2025) provides strong experimental support for this step. This forms the earliest “riboreplisome”, an RNA ribozyme combining replication (“genotype”) with at least one additional self-modification chemistry (“phenotype”). The riboreplisome now possesses all features of a complete genetic system: copied genetic material, copying mechanism, and heritable features subject to selection.

**(4)** Riboreplisomes develop specialisation toward more specific aminoacylation, plausibly demonstrated with self-aminoacylating tRNA ribozymes (Murakami et al., 2003) (**Figure 2c**). As riboreplisomes deplete energised precursors, they evolve capacity to pre-charge ATP (or analogues) with preferred amino acids, a reaction mechanistically similar to 3' end aminoacylation.

**(5)** Combinatorial aminoacylation emerges. Riboreplisome fragments accumulate due to degradation or incomplete replication, eventually becoming charged with amino acids. Using these fragments, or “snatching” from other riboreplisomes, the riboreplisome transfers additional amino acids to its 3' end (**Figure 2d**). As pre-charging becomes specific, interaction specificity develops for discriminated binding of substrates near the catalytic centre based on peptide preferences. This essentially completes a primordial PTC.

**(6)** A duplication or multiplication event occurs as a replication "error" (*e.g.*, slippage producing a covalently-linked copy) (**Figure 3a**, left), or the riboreplisome is *ab initio* a multiplicated arrangement as discussed earlier. This greatly accelerates evolution by relaxing selection criteria and permitting alterations without loss of viability, analogous to how gene duplication enables new function acquisition in modern evolution (Walsh, 1995). Each riboreplisome can now independently charge multiple aminoacyl-AMPs, giving rise to different proto-tRNAs. We note that this duplication is an internal, templated event (slippage producing a covalently linked tandem copy) rather than a 3′-terminal addition, and that the 3′ aminoacylation of Step 3 is reversible and transient (occurring in bind-and-release cycles). The two processes are therefore independent, and 3′ modification does not block duplication. Although lengthening carries a replication cost (proportionally larger for short genomes), the duplicate here is immediately useful rather than redundant. Indeed, a second charging module at once broadens the amino-acid repertoire and hence the quality of replication-enhancing 3′-peptides, so it is selected for its immediate contribution rather than tolerated as a neutral burden (**Table 1**).

**(7)** The "auxiliary" riboreplisome copies rapidly lose redundant replication capacity, retaining only 3' end aminoacylation and amino-acid-charging activities (**Figure 3a**, right). They evolve toward proto-tRNAs that initially function as self-aminoacylating ribozymes with integrated ARS function.

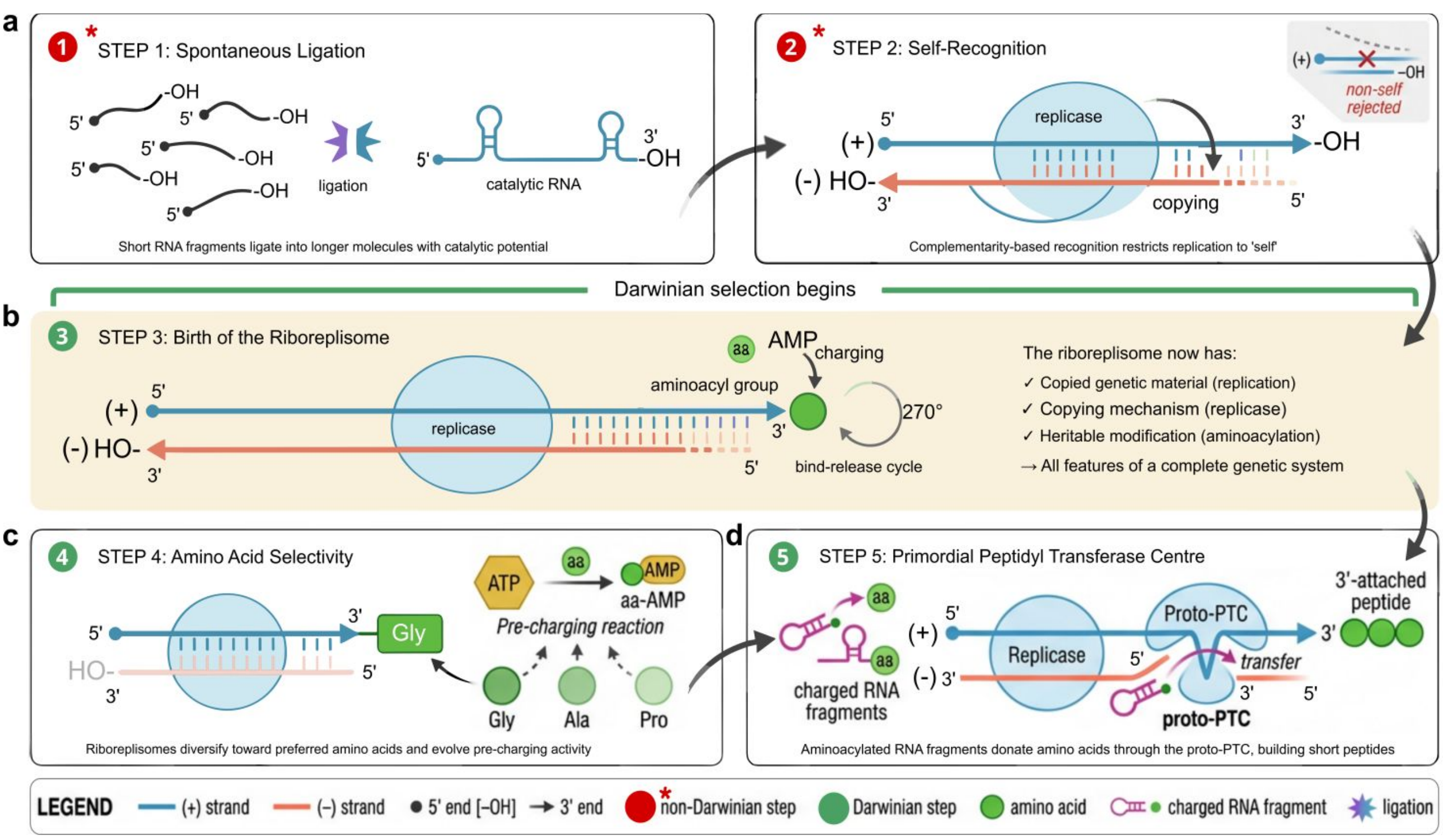

**Figure 2.** Possible pathway of the active centre evolution and function gain by the riboreplisome (Steps 1–5). 12 main steps of the riboreplisome development are depicted across **Figures 2** and **3**, starting from just the replication function and ending up with a system of a developed genetic code that is ready to discard the original riboreplisomal arrangement. RNA termini are depicted as in **Figure 1**. Red asterisks indicate non-Darwinian steps. Purple-and-cyan star indicates ligation. **Step 1:** Short RNA fragments ligate into longer molecules with catalytic potential, generating an RNA replisome ribozyme (blue). **Step 2:** Complementarity-based recognition restricts replication to "self",

establishing a selective framework for retaining subsequent innovations. Steps 1 and 2 are non-Darwinian; from this point, Darwinian natural selection proceeds uninterruptedly. **Step 3:** Self-modification of the 3' end (including aminoacylation from a pre-charged nucleotide available from the environment) occurs as advantageous towards replication efficiency and/or stability, resulting in a first non-polymerisation "feature" and the birth of the riboreplisome, which now possesses all features of a complete genetic system. **Step 4:** Riboreplisomes diversify toward preferred amino acids and evolve pre-charging activity, replenishing aminoacyl-nucleotides from simpler components. **Step 5:** Aminoacylated RNA fragments donate amino acids through the proto-PTC, building short 3'-attached peptides.

**(8)** Facilitated by "gene" multiplication, the riboreplisome develops and expands its internally-encoded codon repertoire for more sophisticated proto-tRNA utilisation and complex peptide synthesis. As these are now genetically encoded and subject to natural selection, co-evolution of the code base and peptides initiates (**Figure 3b**).

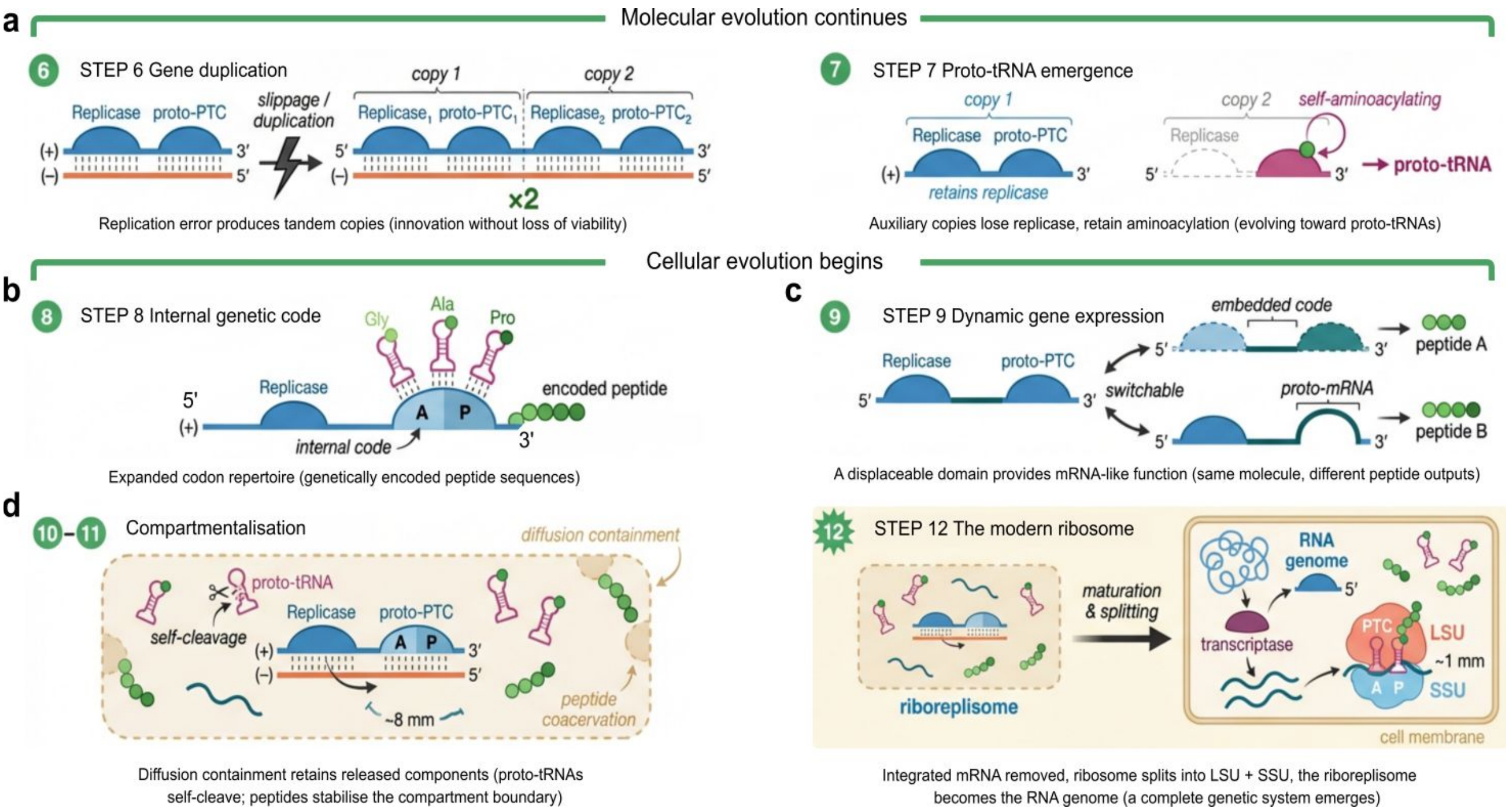

**Figure 3.** Possible pathway of the active centre evolution and function gain by the riboreplisome (continued from **Figure 2**; Steps 6–12). **Step 6:** Replication error (*e.g.*, slippage) produces tandem copies, enabling innovation without loss of viability. **Step 7:** Auxiliary copies rapidly lose replicase activity but retain aminoacylation, evolving toward self-aminoacylating proto-tRNAs. **Step 8:** Expanded codon repertoire enables genetically encoded peptide sequences through co-evolution of the code base and peptides. **Step 9:** A displaceable domain provides mRNA-like function – the same molecule can produce different peptide outputs, lifting the limitation that coding capacity can only change between generations. **Steps 10–11:** Diffusion containment (cell membrane or peptide-based coacervation) retains released components. Proto-tRNAs self-cleave; peptides stabilise the compartment boundary. The riboreplisome's own 3' end modification is replaced with transfer to bound proto-tRNAs, completing the A- and P-site arrangement. **Step 12:** With definitive

compartmentalisation, the integrated mRNA is removed and the proto-ribosome splits into the "decoding" SSU and "catalysing" LSU subunits. The riboreplisome becomes the RNA genome. A complete genetic system emerges.

**(9)** A major limitation remains: ribosomal coding capacity can only change genetically between generations. This limitation is lifted when different riboreplisome fragments, likely arising through multiplication and initial unguided fragmentation, substitute for the internally-coded "message" (**Figure 3c**).

**(10)** At this point, the riboreplisome system develops diffusion restriction and physical containment. Crucially, compartmentalisation is not required to establish genotype-phenotype linkage as it is already provided by the single-molecule architecture (see above), which is precisely why Darwinian selection has operated throughout the preceding steps without any bounding structure. This directly addresses the classical objection that a free, *trans*-acting replicase cannot propagate stably without a compartment: the riboreplisome is not free but linked, so copying acts on "self" and beneficial variation is retained regardless of the unrelated templates in the surroundings. Containment therefore arrives later and for a different reason, namely to capture the now-diffusible released products (aminoacyl-proto-tRNAs and peptides) and raise their local concentration, an enhancement of an already-Darwinian system rather than a precondition for it. Proto-tRNAs evolve self-cleavage and maturation functions, driven by collective advantage of aminoacyl-proto-tRNA availability. Some riboreplisome copies are "sacrificed" for proto-tRNA and proto-mRNA production, and own 3' end modification is replaced with transfer to bound proto-tRNAs, completing the A- and P-site arrangement (**Figure 3d**). The geometry follows directly from this substitution: the peptidyl chain, previously held on the riboreplisome's own modified 3′ end (the precursor of the P-, or peptidyl, site), is now transferred onto an incoming aminoacyl-proto-tRNA bound adjacent to the catalytic centre (the precursor of the A-, or aminoacyl, site). Separating the peptidyl carrier (the riboreplisome 3′ end) from the incoming aminoacyl donor (a discrete proto-tRNA) is what defines the canonical two-site arrangement. Some proto-tRNA and proto-mRNA code may be included in the "negative" riboreplisome strand, beginning separation of gene expression from replication through preferential synthesis of the "negative" strand.

**(11)** With dynamic peptide encoding and environmental release enabling mode-switching even between generations, proto-mRNA code develops to further facilitate encapsulation and coacervation. This may have been the major driver of compartmentalisation and cell-based life (**Figure 3d**).

**(12)** With definitive compartmentalisation, the proto-ribosome becomes permanently programmable by removing (cutting out during maturation) its integrated mRNA sequences and splitting into SSU and LSU rRNA as a result. The unprocessed riboreplisome becomes a "genome" and loses proto-ribosomal identity. The riboreplisome has completed its evolution cycle and transformed into an RNA genome with genetic code, decoding machinery, and gene expression system (**Figure 3d**). This interconnected system is "channelled" toward the modern ribosome because its function now depends vitally on its protein products, including those required for compartmentalisation and cell-like behaviour.

Subsequent development proceeds through well-theorised events: code expansion, translation factor introduction, and replacement of ribozyme-based tRNA pre-charging with aminoacyl-synthetase chemistry. Self-cleavage of RNA genome copies is gradually displaced by gene-selective copying of proto-m-, t-, and rRNA. The emergence of stable cell-like structures eventually enables DNA for information storage, transcribed into RNA on demand. Notably, phylogenetic investigations suggest DNA may have emerged after separation of major cellular branches (Forterre, 2015).

**Table 1. Immediate selective advantage of each step in the riboreplisome pathway.** Steps 1–2 are non-Darwinian bottlenecks (see text); Steps 3–12 each confer a heritable benefit that is selectable in the preceding context, satisfying the requirement that every intermediate stage be advantageous.

| Step | Immediate, heritable selective advantage | Support |
|---|---|---|
| 1–2 | Non-Darwinian bottlenecks: emergence of templated replication and self-restricted copying (prerequisites, not selectable refinements) | Attwater 2018; Gianni 2026 |
| 3 | 3′ self-modification protects the 3′ end from unproductive extension and can activate the replicase via associated residues | Lee 2000; Li 2022 |
| 4 | Specialised pre-charging secures a supply of activated aminoacyl substrate as environmental precursors deplete | Murakami 2003 |
| 5 | Combinatorial transfer builds short 3′-peptides that stabilise and activate the ribozyme (primordial PTC) | Bose 2022; Radakovic 2022 |
| 6 | Duplication relaxes constraint and immediately doubles aminoacylation capacity (more amino-acid types) | Walsh 1995 |
| 7 | Specialisation of auxiliary copies raises aminoacyl supply per unit replication cost | Murakami 2003 |
| 8 | Codon-repertoire expansion enables selectable, genetically encoded peptide sequences | Hartman & Smith 2014 |
| 9 | A displaceable, mRNA-like domain allows peptide output to change within a generation, not only between generations | — |
| 10 | Trans-transfer to bound proto-tRNAs establishes A-/P-site geometry and retains aminoacyl-proto-tRNAs locally | Tamura 2011 |
| 11 | Containment/coacervation retains released catalytic products, raising effective local concentration | Li 2022 |
| 12 | Programmable proto-ribosome decouples coding from replication, enabling unrestricted peptide diversity | — |

To quantitatively evaluate this 12-step pathway, we developed three complementary mathematical models comparing the riboreplisome with alternative “RNA soup” scenarios where multiple independent molecules must cooperate (**Figure 4**).

**Model 1** demonstrates that epistasis creates exponentially rugged fitness landscapes. Using NK fitness landscape modelling (**Figure 5**), the integrated riboreplisome creates a dramatically more navigable evolutionary terrain. In this model, K controls the degree of epistatic interactions between sequence positions. With low epistasis (K=2) reflecting modular genetic linkage, the riboreplisome landscape contains approximately 14-fold fewer local fitness optima compared to multi-component systems with high epistasis, such as the RNA soup (K=8). This ruggedness has severe consequences for evolutionary success: while populations readily find the global optimum on smooth landscapes (K=0, 100% success rate), increasing epistasis rapidly decreases success, with only 5% of populations reaching the global optimum at K=2 and essentially 0% at K≥4 (**Figure 4b**). Crucially, this failure occurs not because evolution stops, as populations continue to adapt and reach similar final fitness values across all K values (**Figure 5**), but because they become trapped in local optima.

## NK Fitness Landscape Model

The NK model, introduced by Kauffman (1993), provides a framework for understanding how epistatic interactions affect the navigability of evolutionary fitness landscapes. In this model, a genome is represented by N binary sites, where each site's contribution to fitness depends on itself plus K other sites (epistatic partners). The parameter K controls landscape ruggedness: K=0 produces a smooth landscape with a single fitness peak, while high K produces rugged landscapes with many local optima that act as evolutionary traps.

We employ the NK model to formalize the intuition that the riboreplisome pathway – where functions evolve sequentially within a single linked molecule – corresponds to evolution on a smooth landscape (low K), while RNA soup scenarios – where multiple RNA species must co-evolve compatible functions – correspond to rugged landscapes (high K).

**Model implementation**:

Genotype space: We use N=12 binary sites, representing the 12 evolutionary steps in the riboreplisome pathway. Each genotype is a binary vector (e.g., `[1,0,1,1,0,0,1,0,1,0,1,1]`) where 1 indicates a step has been achieved and 0 indicates it has not. In this model, steps do not have to be acquired sequentially, capturing all possible scenarios, from the riboreplisome sequential hypothesis to the RNA soup scenario where independent molecules have to co-operate and may acquire steps in any order.

Fitness tables: For each site i, we pre-generate a lookup table containing 2^(K+1) fitness values drawn independently from Uniform(0,1). These values are fixed for the duration of the simulation.

Epistatic neighbors: Each site i is randomly assigned K epistatic partners from among the other N-1 sites. For the riboreplisome model we use K=2 (low epistasis, steps happen sequentially on a linked molecule hence only depend on very few other partners); for the RNA soup model we use K=8 (high epistasis, molecules need to co-operate, hence acquisition of a novel step might only happen in very limited configurations).

Fitness calculation: For a given genotype, the fitness of site i is determined by:

1. Reading the binary states of site i and its K neighbors
2. Converting this (K+1)-bit string to an integer index
3. Looking up the corresponding fitness value in site i's pre-generated table
4. The organism's total fitness is the mean across all N sites

Adaptive walks: Starting from a random genotype, we perform greedy hill-climbing: at each step, we evaluate all N single-bit mutations, identify those that increase fitness, select the best improvement, and repeat until no single-bit mutation improves fitness (a local optimum).

Landscape analysis: For N=12, we exhaustively enumerate all 2^12 = 4096 possible genotypes and identify local optima by checking whether any single-bit mutation increases fitness.

**Key parameters:**

• N = 12 (corresponding to the 12 evolutionary steps in the riboreplisome pathway)
• K_riboreplisome = 2 (low epistasis: each step depends on only 2 other steps)
• K_RNA_soup = 8 (high epistasis: steps are highly interdependent across molecules)
• Number of walks = 50 per landscape (to build confidence intervals on the estimates)
• Number of landscape replicated = 10 per K value (to account for stochasticity in fitness table generation)

**Model outputs**:

For each K value, the model generates:

- Distribution of fitness values across all possible genotypes
- Number of local optima (evolutionary traps)
- Mean final fitness achieved by adaptive walks from random starting points
- Steps to local optimum (evolutionary trajectory length)

**Key result (Figure 5)**: Riboreplisome landscapes (K=2) exhibit 10.8± 3.9 local optima with mean final fitness >0.71, while RNA soup landscapes (K=8) exhibit 155.4 ± 11.4 local optima with mean final fitness <0.68, demonstrating that coordination requirements create evolutionary obstacles.

**Figure 5.** NK Fitness Landscape Analysis. Comparison of fitness landscapes under the riboreplisome model (low epistasis, K=2) versus RNA soup model (high epistasis, K=8). N = 12 sites, 50 walks per landscape, 10 landscape replicates per K value. **(a)** Number of local optima (evolutionary traps) increases exponentially with epistasis K, demonstrating that multi-component coordination creates a dramatically more rugged fitness landscape. **(b)** Probability of reaching the global optimum in a single adaptive walk from a random starting point. The riboreplisome model (K=2) retains approximately 30% probability, while the RNA soup model (K=8) has near-zero probability. **(c)** Mean final fitness achieved after adaptive walks. Despite achieving similar fitness values across K, populations at high K are trapped in local optima rather than reaching the global optimum. **(d)** Mean number of steps before reaching a local optimum. The non-monotonic pattern reflects that at very low K, landscapes are smooth and walks reach the (global) optimum quickly; at intermediate K, moderate ruggedness allows longer exploratory walks; at high K, populations become trapped almost immediately in nearby local optima, paradoxically shortening path length.

Model 2 reveals that riboreplisome architecture enables sequential information accumulation while RNA soup requires simultaneous multi-step innovations. Information-theoretic analysis (**Figure 6**) shows that only 39% of total functional information (65 of 165 bits) requires non-Darwinian acquisition (**Figure 6**), concentrated entirely in Steps 1-2. The remaining 10 steps accumulate information through selection. Transition probability analysis (**Figure 4c**) reveals the mechanistic basis: the riboreplisome maintains forward progress through selection (diagonal probability matrix showing each k-step state is stable), with Darwinian steps proceeding at probabilities approximately five orders of magnitude higher than the initial bottleneck steps. In contrast, RNA soup uses an all-or-nothing strategy where states with partial information revert to zero unless all k steps co-occur. Consequently, >99.99% of total evolutionary waiting time concentrates in just 2 of 12 transitions. Once the initial RNA replicase emerges and acquires 3'-modification capacity, subsequent riboreplisome evolution proceeds rapidly and inevitably (**Figure 6c**).

## Information Accumulation Model

This model employs Adami's physical complexity framework (Adami, 2002), which defines biological complexity as the mutual information between a genome and its environment accumulated through natural selection. Physical complexity is measured in bits and can only increase incrementally through Darwinian evolution, as each selectively retained mutation adds functional information relevant to survival and reproduction.

We apply this framework to quantify the informational requirements of each step in the 12-step riboreplisome pathway and compare two assembly scenarios: (1) incremental evolution where each step is selected before the next emerges, versus (2) random assembly where all components must arise simultaneously.

Information content per step was estimated based on the functional requirements of each evolutionary transition:

• Step 1 (RNA replicase emergence): 40 bits –
• Step 2 (3' modification): 25 bits –
• Steps 3-12 (Darwinian): 8-12 bits each – incremental refinements under selection

Total information: 165 bits
Non-Darwinian contribution (Steps 1-2): 65 bits (39.4%)
Darwinian contribution (Steps 3-12): 100 bits (60.6%)

The model compares incremental evolution (riboreplisome) where information accumulates step-by-step with selection at each stage, versus random assembly (RNA soup) where all components must arise independently and find each other.

**Model implementation**:

<u>Information budget assignment</u>: Each of the 12 steps is assigned an information content (bits) based on:

- Functional complexity (number of specific nucleotides required)
- Literature precedents (e.g., Bartel and Szostak, 1993 selected functional ribozymes from $10^{15}$ sequences, corresponding to ~40 bits)
- Whether the step builds incrementally on previous functions (Darwinian) or requires a novel catalytic activity (non-Darwinian)

Step classifications:

- Steps 1-2: Non-Darwinian (40 and 25 bits respectively) - major innovations with no selective intermediate. Step 1 requires catalytic core, substrate binding. Step 2 requires self-recognition, aminoacylation chemistry.
- Steps 3-12: Darwinian (8-12 bits each) - refinements and elaborations with selectable intermediates

Incremental evolution simulation: For each step i:

1. Draw waiting time from geometric distribution with probability p_i: wait_time ~ Geom(p_i)
2. Accumulate generation count: total_gen = total_gen + wait_time
3. Upon achieving step i, add its information content to cumulative total
4. Proceed to step i+1 with selection maintaining progress

Random assembly simulation (all-or-nothing dynamics):

1. At each generation, each of the 12 steps occurs independently with probability p_i
2. Count number of steps k that co-occur in that generation
3. Calculate transient information = sum of bits from those k steps
4. **Critical constraint**: If k < 12, information is immediately lost (no selection for partial function). Only if k = 12 do all steps occur simultaneously → success (information retained)

Probability calculations:

- Expected waiting times
  - Non-Darwinian steps (1-2): ~$10^{10}$ generations total
  - Darwinian steps (3-12): ~$10^{3}$ generations total
- Expected generations for incremental pathway: Σ(1/p_i) ≈ 10^11 generations
- Probability of k-step coincidence in random assembly: maximum over all k-combinations of ∏(p_i for i in combination)
- Probability of 12-step coincidence: prod(p_1, p_2, ..., p_12) ≈ 10^-60 per generation
- Trials needed for 50% probability of random assembly: ln(0.5)/ln(1 - P_all) ≈ 10^60 generations
- Advantage ratio: ≈10^49

**Results (Figure 6):**

The incremental pathway successfully accumulates all 165 bits of functional information, crossing the minimal function threshold of 100 bits. In contrast, the random assembly model fluctuates randomly around ~15 bits (11% of target) and never achieves functional status. The probability ratio favoring incremental evolution is effectively infinite (P_random ≈ $10^{-37}$ per trial). The key finding is that >99.99% of the total evolutionary waiting time is concentrated in just 2 of the 12 steps. Once the initial non-Darwinian bottlenecks are crossed (RNA replicase emergence and 3' modification capacity), the remaining 10 steps proceed rapidly *via* standard Darwinian selection. **ure 6**. Information accumulation analysis comparing incremental evolution (riboreplisome) versus random assembly (RNA soup) models. **(a)** Information content (in bits) required for each of the 12 evolutionary steps. Steps 1–2 (red) are non-Darwinian bottlenecks requiring major innovations (65 bits total, 39% of pathway). Steps 3–12 (blue) are Darwinian refinements with selectable intermediates (100 bits total). Total pathway requires 165 bits of functional information. **(b)** Per-generation transition probability matrices ($\log_{10}$ scale). States represent the number of steps achieved (0 to 12). **Riboreplisome model** (left): Selection maintains progress – from state *k*, can only stay at *k* or advance to *k*+1, but steps are ordered. State 12 is absorbing (functional riboreplisome achieved). **RNA soup model** (right): All-or-nothing dynamics – partial progress (states 1–11) is non-functional and immediately lost (returns to state 0), but any step may occur first. From state 0, achieving *k* steps simultaneously requires the *k* most favorable steps to co-occur (shown as maximum probability among all *k*-step combinations). Only state 12 is stable. **(c)** Information accumulation dynamics over evolutionary time. Green step function shows a representative riboreplisome simulation: horizontal plateaus represent waiting times for each step, with vertical jumps upon acquisition. Information accumulates monotonically and irreversibly through selection. Colored vertical spikes show RNA soup transient information gains (spikes colored by number of steps co-occurring): partial combinations arise randomly but collapse immediately without complete function, never building on prior progress. **(d)** Frequency distribution of multi-step coincidences in the RNA soup model. Boxplots show variation across 10 independent replicate simulations (50 simulations per replicate, $10^{11}$ generations each). Single-step events are common; multi-step coincidences become exponentially rarer. Twelve-step coincidence (required for function) never observed in 500 simulation runs totaling $5\times10^{13}$ generation-equivalents. **(e)** Distribution of maximum information achieved across simulations after $10^{11}$ generations. Riboreplisome simulations (green) all reach the 165-bit target. RNA soup simulations (red) achieve at most ~40 bits (23% of target) through transient coincidences, never approaching functional threshold. Dashed lines indicate mean values.

Model 3 quantifies how diffusion loss and architectural constraints limit multi-component RNA systems (**Figure 4d**; **Figure 7**). Direct simulation of single-molecule versus multi-molecule evolutionary dynamics demonstrates that the RNA soup model suffers fitness decline primarily due to diffusion loss of catalytic products away from the replicating genome. Systematically removing each RNA soup problem reveals that diffusion is the dominant factor: eliminating diffusion increases RNA soup fitness 2.76-fold to ~1.3, recovering most of the deficit relative to the riboreplisome (**Figure 4e**). In contrast, removing cheater emergence provides only 1.04-fold improvement and removing linkage loss has no effect (0.93-fold), indicating that while these problems are theoretically plausible, they are negligible under realistic parameter conditions. The riboreplisome achieves 5.69-fold higher fitness than baseline RNA soup, stemming from both the elimination of diffusion loss (~2.76-fold) and additional architectural benefits (~2-fold multiplicative advantage from inherent single-molecule constraints on multi-component coordination). Parameter sensitivity sweeps confirm that RNA soup fitness is highly sensitive to diffusion rate but shows minimal response to cheater emergence or linkage loss rates. This quantitative framework provides strong support for the riboreplisome as the evolutionarily viable pathway from prebiotic chemistry to cellular life.

## Single vs Multi-Molecule Comparison

This model directly compares the evolutionary dynamics of single-molecule (riboreplisome) versus multi-molecule (RNA soup) architectures, explicitly incorporating cheater dynamics that arise in multi-component systems.

**Model implementation**:

Parameters: set of universal parameters:

- Population size: 500 individuals
- Generations: 500
- Mutation rate: 1% per generation, out of which
    - Beneficial mutations: 2% probability, +15% fitness effect
    - Deleterious mutations: 5% probability, -5% fitness effect

Riboreplisome-specific:

- Diffusion loss: 0% (genetic linkage keeps products associated)
- Linkage loss: 0% (components cannot separate)
- Cheater emergence: 0% (single molecule cannot cheat itself)

RNA soup-specific:

- Diffusion loss: 0.2% per generation (products diffuse away from genes), and sensitivity analysis from 0 to 0.5%
- Linkage loss: 25% per generation (combinations break apart) 0% to 100% per generation, and sensitivity analysis by increment of 0.25 per simulation batch
- Cheater emergence: 1% per component per generation (selfish replicators), and sensitivity analysis from 0% to 5%
- Coordination requirement: All 4 components must be functional for cooperation

Implementation:

Each simulation tracks 500 individuals over 500 generations using discrete-generation Wright-Fisher dynamics. Every individual experiences stochastic mutation, selection, and model-specific processes (linkage loss, diffusion, cheater emergence).

- Mutation System

Each generation, individuals have a 1% probability of mutation affecting the entire integrated molecule. Mutations are beneficial (2% probability, +15% fitness effect), deleterious (5% probability, -5% fitness effect), or neutral (93% probability). This yields a net expected fitness change of +0.3% - 0.25% = +0.05% per generation, allowing evolutionary improvement.

- Linkage Loss (RNA soup only)

During each generation, component associations can break and reform. Each individual has a probability (0-100% tested) of swapping one randomly-selected component with another random individual in the population.

- Diffusion Loss (RNA soup only)

Catalytic products diffuse away from replicating molecules, reducing local substrate concentration and replication efficiency. This is modeled as a constant multiplicative fitness reduction each generation. 0% to 0.5% fitness loss per generation. Products remain tethered near the active site due to genetic linkage between catalytic components and the replicating genome.

- Cheater Emergence (RNA soup only)

In cooperative multi-component systems, selfish variants can evolve that replicate without contributing to group function. Each component has a per-generation probability (0-5% tested) of becoming a "cheater" that replicates 20% faster but contributes zero value to the fitness calculation.

RNA soup fitness is calculated as the geometric mean of component values. If any component becomes a cheater (value = 0), the geometric mean approaches zero, making the individual effectively inviable. Selection rapidly removes cheater-infected individuals unless they are rescued by recombination (linkage loss swapping the cheater away) or hitchhiking with exceptionally high-quality other components.

- Coordination Bottleneck (RNA soup only)

In some scenarios, we tested whether RNA soup faces a coordination requirement: all four components must be of similar quality to avoid fitness penalties. Specifically, if the weakest component falls below 70% of the strongest component's value, the individual suffers a 50% fitness penalty.

<u>Fitness Calculation and Selection:</u>

**Riboreplisome**: Fitness is a single real number tracking the integrated molecule's replication efficiency.

**RNA soup**: Fitness is the geometric mean of the four component values. Geometric mean (rather than arithmetic mean) is used because components are multiplicative - one failed component (value approaching zero) renders the entire system non-functional. The geometric mean of [a, b, c, d] is $(a\times b\times c\times d)^{(1/4)}$.

Selection follows fitness-proportional sampling (Wright-Fisher model): each individual's probability of contributing offspring to the next generation equals its fitness divided by total population fitness. The next generation is sampled with replacement according to these probabilities.

**Results (Figure 7):**

The riboreplisome model shows sustained fitness growth from 1.0 to ~2.61 over 500 generations. In stark contrast, the RNA soup model collapses no matter the combination of parameters. Genetic linkage provides riboreplisome advantage primarily by preventing product diffusion (~33% of effect), with fundamental architectural constraints (multi-component coordination, mutation limitation) contributing the majority (~66%). Social cheating and linkage loss, while biologically plausible, are minor factors under these conditions.

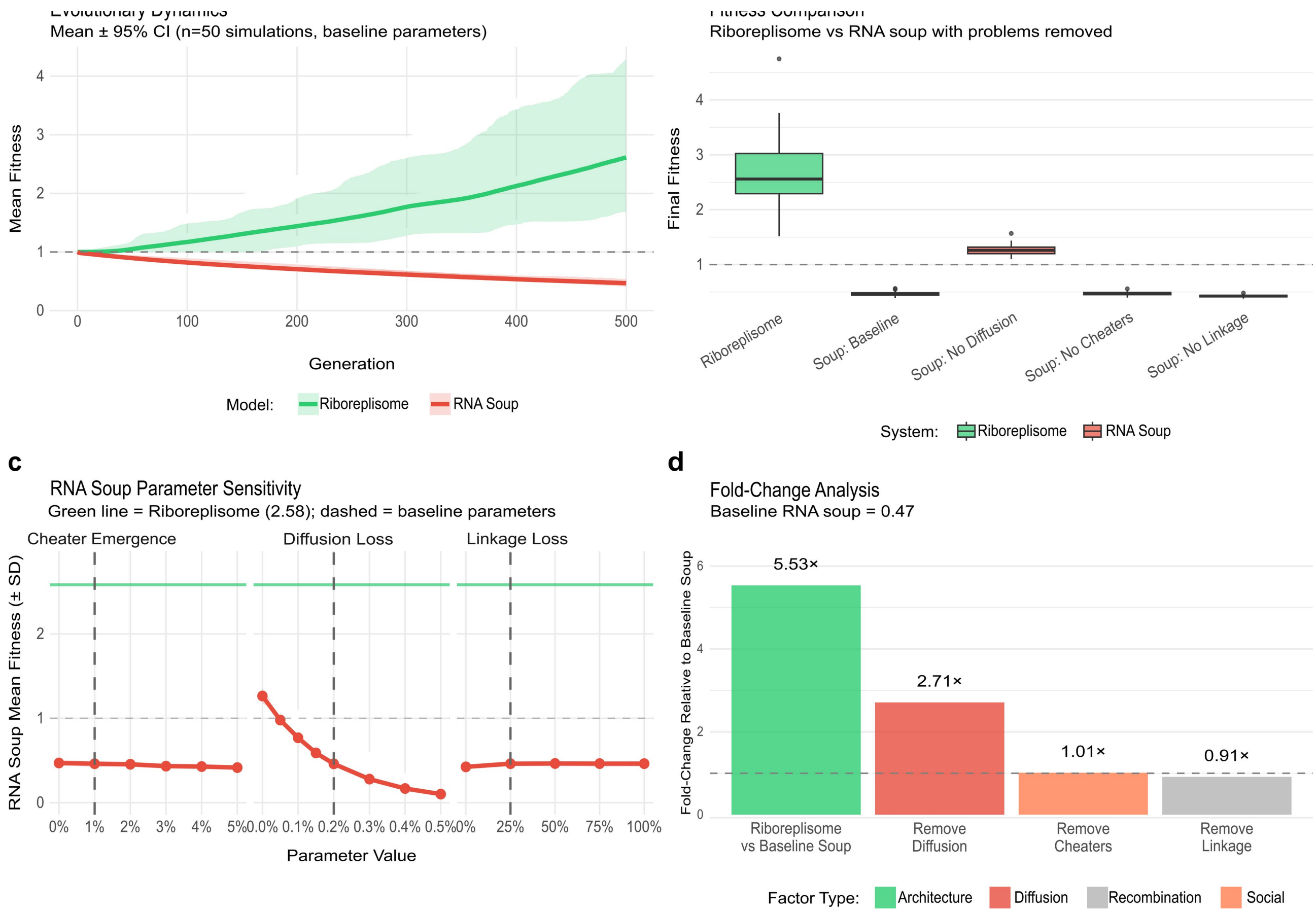

**Figure 7.** Riboreplisome advantage over RNA soup stems from differential diffusion and architectural constraints. **(a)** Evolutionary trajectories. Mean fitness (±95% CI) over 500 generations for riboreplisome (green) and RNA soup (red) under baseline conditions (RNA soup: 0.2% diffusion, 1% cheater emergence, 25% linkage loss; both systems: 1% mutation rate). Riboreplisome fitness increases steadily (final: 2.61 ± 0.66, n=50 simulations) while RNA soup declines (final: 0.46 ± 0.02). Dashed line indicates initial fitness (1.0). **(b)** Fitness across scenarios. Boxplots show final fitness distributions (n=100 simulations each) comparing riboreplisome to RNA soup with individual problems removed. Removing diffusion restores RNA soup fitness to ~1.3 (2.76× improvement over baseline), demonstrating diffusion is the primary limiting factor. Removing cheaters provides modest improvement (1.04×), while removing linkage loss has no effect (0.93×). **(c)** Parameter sensitivity analysis. RNA soup fitness response to varying diffusion rate (0-0.5%), cheater emergence rate (0-5%), and linkage loss rate (0-100%). Green horizontal line shows stable riboreplisome fitness (2.61 ± 0.66). Dashed vertical lines mark baseline parameters. RNA soup is highly sensitive to diffusion rate (strong negative slope) but shows minimal response to cheater emergence and no response to linkage loss. Point at 0% diffusion shows RNA soup can thrive when diffusion is eliminated, while still not reaching Riboreplisome fitness. **(d)** Decomposition of fitness advantage. Contributions of each factor to fitness differences, expressed as fold-change relative to baseline RNA soup (0.46). Riboreplisome achieves 5.69× higher fitness. Removing diffusion provides 2.76× improvement (largest single factor for RNA soup). Removing cheaters provides minimal improvement (1.04×), and removing linkage loss has no effect (0.93×). The riboreplisome's advantage stems from both eliminating diffusion (same

as RNA soup without diffusion would achieve, ~2.76×) plus additional architectural benefits (~2× multiplicative advantage).

## Parameter Justification

All model parameters were derived from published experimental and theoretical studies. Key parameters and their justifications are provided in **Table 2**.

**Table 2.** Model Parameters and Justifications

| Parameter | Value | Model(s) | Justification/Reference |
|---|---|---|---|
| N (evolutionary steps) | 12 | 1, 2 | Manuscript Figures 2–3 |
| K_ribореplisome | 2 | 1 | Low epistasis for sequential, modular evolution |
| K_RNA_soup | 8 | 1 | High epistasis for multi-component coordination |
| Landscape replicates | 10 per K | 1 | Stochasticity in fitness table generation |
| Walks per landscape | 50 | 1 | Confidence intervals on trajectory outcomes |
| I_threshold (minimal) | 100 bits | 2 | Estimated minimal functional information |
| I_threshold (full) | 165 bits | 2 | Sum of all 12 steps (see below) |
| Step 1 information | 40 bits | 2 | Bartel and Szostak (1993) ribozyme selection from $10^{15}$ sequences |
| Step 2 information | 25 bits | 2 | Lee et al. (2000) aminoacylation ribozyme |
| Darwinian step info | 8–12 bits | 2 | Incremental refinement under selection |
| P_step1 (replicase) | $10^{-10}$ | 2 | Rare emergence from random pool |
| P_step2 (3' mod) | $10^{-7}$ | 2 | Builds on existing function |
| P_Darwinian | $10^{-2}$ | 2 | Readily achievable with selection |
| Population size | 500 | 3 | Wright-Fisher dynamics; sensitivity not critical |
| Generations | 500 | 3 | Sufficient to observe fitness trajectories to equilibrium |
| Simulations | 100 | 3 | Statistical confidence (reported as n=50 or n=100) |
| Mutation rate | 1% per generation | 3 | Typical RNA replication error rate (Eigen, 1971) |
| Beneficial mutation | 2% of mutations | 3 | Conservative estimate |
| probability | | | |
| Beneficial effect | +15% fitness | 3 | Strong selection (>> drift barrier) |
| Deleterious mutation probability | 5% of mutations | 3 | Moderate purifying selection |
| Deleterious effect | −5% fitness | 3 | Asymmetric (beneficial > deleterious in magnitude) |
| Diffusion loss (RNA soup baseline) | 0.2% per generation | 3 | Products diffuse away from replicating genome |
| Linkage loss (RNA soup baseline) | 25% per generation | 3 | Component associations break and reform |
| Cheater emergence (RNA soup baseline) | 1% per component per generation | 3 | Selfish replicator variants |
| Number of RNA soup components | 4 | 3 | Proto-replicase, proto-tRNA, proto-rRNA, proto-mRNA |

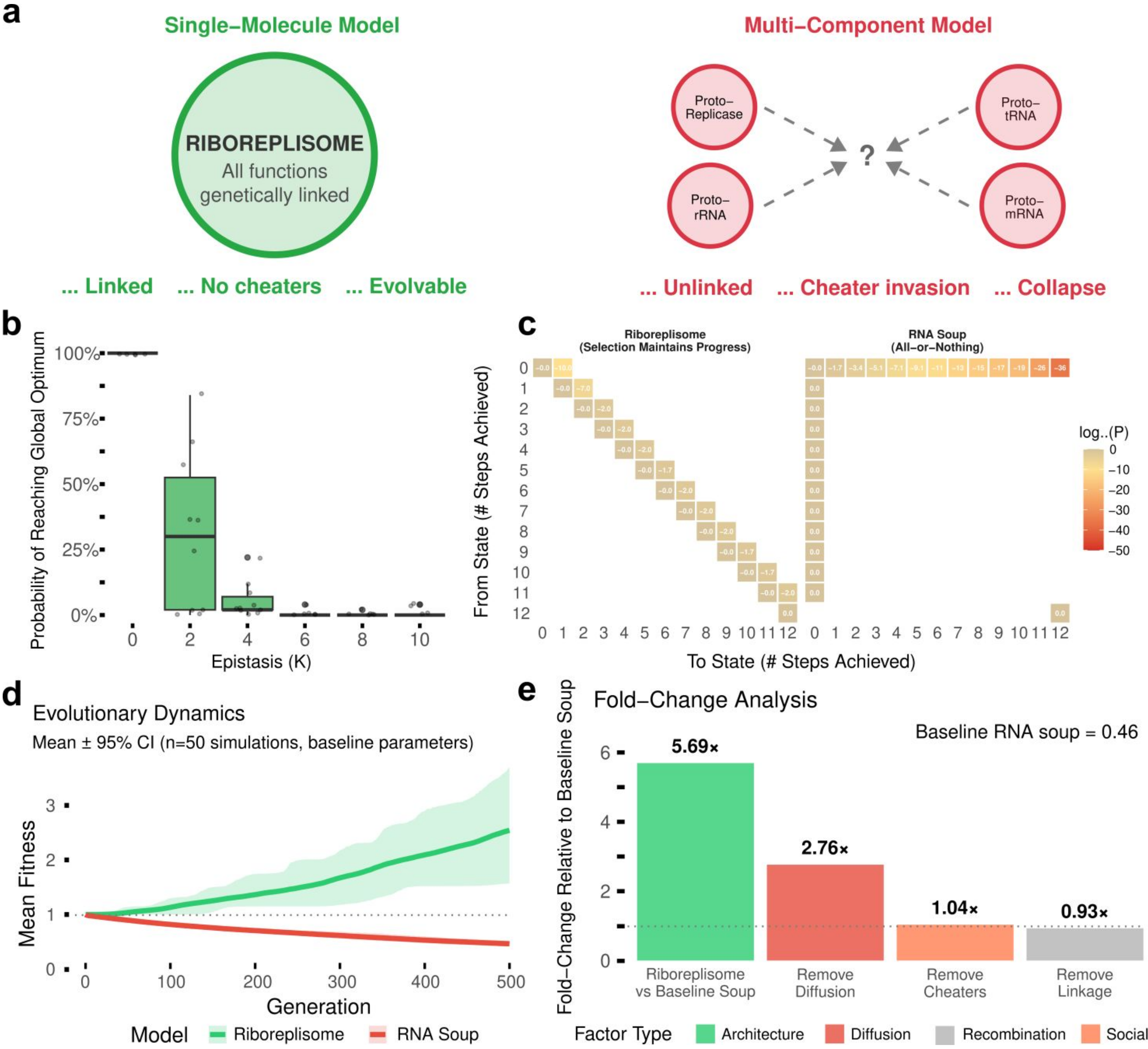

**Figure 4.** Single-molecule architecture provides evolutionary advantage over multi-component models. **(a)** Conceptual comparison of the riboreplisome model, where all functions are genetically linked within a single RNA molecule, versus multi-component "RNA soup" models requiring cooperation between independent molecules. **(b)** Probability of reaching global optimum in a single trajectory from a random starting point. The riboreplisome model has a 30% chance of reaching optimum, while the probability is almost null for the RNA soup model. **(c)** Per-generation transition probability matrices (log10 scale). States represent the number of steps achieved (0 to 12). Riboreplisome model (left): Selection maintains progress: from state k, can only stay at k or advance to k+1, but steps are ordered. State 12 is absorbing (functional riboreplisome achieved). RNA soup model (right): all-or-nothing dynamics as partial progress (states 1–11) is non-functional and immediately lost (returns to state 0), but any step may occur first. From state 0, achieving k steps simultaneously requires the k most favorable steps to co-occur (shown as maximum probability among all k-step combinations). Only state 12 is stable. **(d)** Simulated fitness trajectories over 500

generations (n=100 simulations per model; shaded region indicates 10th-90th percentile). The riboreplisome shows sustained fitness increase while RNA soup fitness declines, primarily due to diffusion loss of catalytic products. **(e)** Contributions of each factor to fitness differences, expressed as fold-change relative to baseline RNA soup. Riboreplisome achieves 5.69× higher fitness. Removing diffusion provides 2.76× improvement (largest single factor), while removing cheaters (1.04×) or linkage loss (0.93×) has minimal effect. The riboreplisome's advantage stems from eliminating diffusion (~2.76×) plus additional architectural benefits (~2× multiplicative advantage).

## Discussion

The catalytic RNA core of the ribosomal decoding centre, despite being relictual to the extreme, preserves designs of the original universal RNA machine. The riboreplisome concept provides an uninterrupted link between chemical and biological evolution, a pathway for the emergence of the ribosome and, consequently, cellular life.

### *Molecular relics of the riboreplisome*

Several features arising from purely theoretical considerations bear remarkable resemblance to current ribosomal organisation. Contemporary ribosomes are predominantly transcribed as single pre-rRNA units containing embedded tRNAs, with ribosome production tightly coordinated through the biogenesis “regulon” (Brown et al., 2008; Henras et al., 2015). Transfer-messenger RNA (tmRNA) evidences natural consolidation toward singular RNA molecules combining translation functions (Macé and Gillet, 2016; Withey and Friedman, 2003). Evidence suggests SSU rRNA arose through concatenation of proto-tRNA molecules (de Farias et al., 2021), consistent with proposed riboreplisome multiplication. Fragments resembling tRNAs and ORFs encoding key proteins can be reconstructed within rRNA sequences, indicating possible common ancestry (Root-Bernstein and Root-Bernstein, 2015, 2019). The “circular code” hypothesis, which postulates a reduced early code with self-complementary sequences located in structural rRNA regions, does provide a converging support (Dila et al., 2019).

Regarding ribosome dynamics, minihelix interactions with tRNA-like function can explain initial PTC operation (Tamura, 2011). The CCA stem represents the more ancient part of tRNAs, with the anticodon loop showing signs of later evolution (Di Giulio, 2009; Maizels and Weiner, 1994). The evolutionary-older h44 and h28 of bacterial SSU link to inter-subunit movement that becomes ratcheting motion (Fox et al., 2015). Evidence that EF-Tu evolved before EF-G suggests the riboreplisome initially possessed only a single functional site (Hartman and Smith, 2010). Resurrected ancestral EF-Tu variants show reduced accuracy but increased promiscuity, and surprisingly remain active with modern ribosomes (De Tarafder et al., 2021).

The hypermodification of rRNA deserves attention. tRNAs and rRNA contain exceptionally high modified nucleotide content, concentrated near functional centres (Decatur and Fournier, 2002; Sloan et al., 2017). These modifications may represent “scars” of rRNA evolution, alternative chemical diversity developing parallel to protein evolution. That unmodified SSU rRNA can evolve high activity after just 15 selection cycles demonstrates the flexibility underlying this system (Murase et al., 2018). RNA modification enzymes may encode some of the most ancient protein functions.

***Selective advantages of the riboreplisome***

What drove proto-ribosomal function? Any heritable feature enhancing propagation would confer enormous benefit. Two categories emerge: (1) direct replication enhancement through partner binding, condensate formation, or ribozyme activation *via* positively charged residues; and (2) preservation through 3' end protection against parasitic extension, which likely was the most immediate threat given riboreplisome self-replication properties. Hydrophobic cationic peptides substantially enhance RNA polymerase ribozyme activity (Li et al., 2022), suggesting the aboriginal proteome comprised peptides improving riboreplisome function. Initial genetic selection may have occurred cell-free, with successful peptides remaining associated with their genetic carriers through direct binding, which had been experimentally implemented in ribosome display (Hanes and Plückthun, 1997).

The single-molecule architecture's superiority over multi-component "RNA soup" models warrants emphasis. Multi-component systems face fundamental challenges: catalytic products diffuse away from the replicating genome, molecules exploiting shared resources without contributing can gain selective advantage as "cheaters", and cooperative molecular assemblies are fragile to dissociation. Our simulations demonstrate that diffusion loss is the dominant limiting factor, with eliminating diffusion alone recovering most of the RNA soup's fitness deficit (2.76-fold improvement). Cheater invasion and linkage loss, while theoretically concerning, prove negligible under realistic conditions. The riboreplisome circumvents all of these problems through genetic linkage where beneficial mutations in any domain co-replicate with all others, catalytic products remain tethered near the active site, and the single-molecule architecture is inherently immune to cooperation collapse. The resulting ~6-fold fitness advantage likely underestimates the true benefit, as the architectural constraints of multi-component coordination impose additional limitations beyond those captured by any single parameter.

***Genetic code evolution and ribosome miniaturisation***

The riboreplisome offers an uninterrupted path for genetic code sophistication through proto-tRNA alteration and codon expansion. All code variants likely existed but were outcompeted; deep thermodynamic and combinatorial reasons underlie current code success (Grosjean and Westhof, 2016; Westhof et al., 2022). From this perspective, the code never appears randomly coincidental but always represents sequential sophistication from simple initial states. Early cellular life with alternative codes may have existed but became extinct through code sub-optimality.

The ribosome represents an ongoing miniaturisation work: larger RNA building blocks are progressively replaced by smaller, more diverse proteins (Bernhardt and Tate, 2015). Mitochondrial ribosomes exemplify this trend, with 19 new proteins added since endosymbiosis while rRNA content decreased (Desmond et al., 2011). Paradoxically, organisms with complex gene expression possess longer rRNA (Bowman et al., 2015), suggesting rRNA provides essential functional continuity preventing excessive "channelling" toward irreversible optimisation. The riboreplisome was – and the ribosome remains – an irreplaceable evolutionary vehicle: sufficiently crude not to be easily broken, yet accurate, forgiving, robust, and modular.

***Quantitative support and testable predictions***

The riboreplisome hypothesis assumes just two non-Darwinian steps: RNA-dependent RNA polymerase emergence and acquisition of 3' end-modifying function. Both are experimentally supported: the first now compellingly demonstrated by the discovery of a 45-nucleotide polymerase ribozyme capable of synthesising both itself and its complementary strand from random sequence pools (Gianni et al., 2026), the second partially confirmed through ribozymes capable of 3' end operations (Lee et al., 2000; Murakami et al., 2003). Recent demonstration of spontaneous RNA aminoacylation *via* thioesters under prebiotic conditions (Singh et al., 2025) provides strong support for the self-aminoacylation mechanism. Our quantitative modelling demonstrates these bottleneck steps (transition probabilities ~$10^{-10}$ and ~$10^{-7}$) account for essentially all evolutionary difficulty; the subsequent ten Darwinian steps proceed at ~$10^{-2}$, representing five to eight orders of magnitude difference (**Figure 6**). Only 39% of total functional information (65 of 165 bits) requires non-Darwinian acquisition (**Figure 5**).

The hypothesis generates several testable predictions:

(1) Synthetic reconstruction of minimal riboreplisomes combining replication and aminoacylation activities should be achievable through directed evolution;

(2) Riboreplisome-like sequences or structural motifs may be discoverable within rRNA through phylogenetic reconstruction;

(3) Single-molecule translation systems should demonstrate cheater resistance compared to multi-component alternatives in laboratory evolution experiments;

(4) The 12-step pathway predicts specific intermediate structures that could be reconstructed and tested for function.

The recent structural elucidation of RNA polymerase ribozymes (McRae et al., 2024) demonstrates that complex, multi-domain RNA machines capable of template-directed synthesis are structurally achievable, and, most remarkably, the discovery of a 45-nucleotide polymerase ribozyme that can synthesise itself and its complementary strand (Gianni et al., 2026), together demonstrate that the core catalytic capabilities assumed by the riboreplisome hypothesis are experimentally achievable in compact RNA motifs. That such polymerase activity was found in a motif approaching lengths accessible by nonenzymatic RNA polymerisation suggests that the emergence of replicase ribozymes from prebiotic chemistry may be far more probable than previously estimated.

***Conclusion***

The riboreplisome offers answers to the fundamental question of what constitutes the smallest “living” entity. It enables Darwinian evolution earlier than RNA-protein co-evolution models require, explains seamless sophistication of programmed features without requiring compartmentalisation, yet provides driving force toward cellular organisation once peptide release emerges. The riboreplisome explains origins of all three universal RNA types – rRNA, tRNA, and mRNA – and provides basis for RNA processing, maturation, and eventually transcription. While the aboriginal riboreplisome is unlikely recoverable, its relics persist across biological RNA. “Reverse engineering” a riboreplisome would provide synthetic biology opportunities in self-contained translation/replication systems built on parallel genetic codes, and represent a first step in replicating life’s emergence.

**Acknowledgements**

The authors are very grateful for the productive and very interesting discussions around the topic, and major insight and advice, to John Mattick and Anna Wang. The authors would like to acknowledge their use of the Kaya HPC facility at UWA and the unwavering, highly enabling support provided by the Kaya HPC team, including the amazing contributions and work of Chris Bording.

**Author contributions**

N.E.S., conception and initial draft and mathematical modelling, A.C., conception and draft and model refinement, A.H.F. manuscript refinement and conceptual model input, all authors have edited the manuscript and prepared the figures.

**Funding**

European Union’s Horizon 2020 – Research and innovation program Marie Sklodowska-Curie grant (agreement No 890462) to A.C.; National Health and Medical Research Council of Australia (NHMRC) Investigator Grant (GNT1175388) to N.E.S.; Bootes Foundation Grant (2022) to N.E.S.; Australian Research Council Discovery Grant (DP180100111, DP250103133) to N.E.S.; CNRS Postes Rouges Grant (2025) to N.E.S.; National Health and Medical Research Council of Australia Grant (APP1147496) to A.H.F.; Australian Research Council Grant (FT180100204) to A.H.F.